# A high performance membraneless microfluidic microbial fuel cell featuring stable, long-term operation under a wide range of flow rates

Mehran Abbaszadeh Amirdehi,[a] Nastaran Khodaparastasgarabad,[a] Hamza Landari,[b] Mir Pouyan Zarabadi,[a] Amine Miled,[b] and Jesse Greener *[a,c]

---

[a]  M. A. Amirdehi, N. Khodaparastasgarabad, M.P. Zarabadi, Prof. J. Greener
     Département de Chimie, Université Laval.
     2325 Rue de l'Université, Québec, Canada.
     E-mail: Jesse.Greener@chm.ulaval.ca
[b]  H. Landri, Prof, A. Miled
     Département de Génie électrique, Université Laval.
     2325 Rue de l'Université, Québec, Canada.
[c]  Prof. J. Greener
     CHU de Québec, centre de recherche, Université Laval.
     10 rue de l'Espinay, Québec, QC, Canada.

**Abstract:** Strong control over experimental conditions in microfluidic channels provides a unique opportunity to study and optimize membraneless microbial fuel cells (MFCs), particularly with respect to the role of flow. However, improved performance and transferability of results to the wider MFC community require improvements to device operational times, oxygen shielding, and anode/cathode compartment isolation. To address these challenges, we present an easy-to-fabricate membraneless MFC with integrated graphite electrodes and an oxygen protection barrier. Guided by computational fluid dynamics, the device architecture was optimized to account for the presence of the anode-adhered electroactive biofilm to prevent cross-contamination between the anode and cathode compartments under a wide range of flow rates. Attention to all of these design features in the same platform resulted in operation of a MFC with a pure-culture anaerobic *Geobacter sulfurreducens* biofilm for half a year, six times longer than previously reported, without the use of an oxygen scavenger. As a result of higher device stability under high flow rates, power densities were four times higher than reported previously for microfluidic MFCs with the same biofilm.

## Introduction

Microfluidics has emerged as an essential technology in microbiological research, primarily due to its ability to provide rigorous control over experimental conditions.[1-4] A potential area for expanded contribution lies in bio-electrochemical systems (BES), an expanding subdomain of synthetic biology, in which biocatalytic redox steps in certain microbial metabolic cycles perform useful electrochemical functions.[5,6] Electroactive biofilms (EABs) from anaerobic *Geobacter sulfurreducens* are among the most studied because of their special ability to directly transport electrons to an anode via a conductive extracellular network consisting of cytochrome c proteins and conductive pili.[7] Compared with mixed cultures, the relative simplicity of monoculture EABs from *G. sulfurreducens* makes it easier to gain mechanistic insights. For example, using three-electrode microfluidic BES devices with pure *G. sulfurreducens* electrode-adhered EABs, we recently studied the role of liquid flow on the kinetics of electrobiocatalysis,[8] EAB pH management,[9] and bacterial metabolic state under highly nutrient-limited conditions.[10] Additionally, accurate simulations of laminar flow streams in microchannels could be achieved with straight-forward models.[11,12] The most highly developed microfluidic BES are microbial fuel cells (MFCs), which are devices with anode-adhered EABs that convert chemical potential energy from organic waste molecules into electrical energy via biocatalytic oxidation.[13,14] Among other advantages, microfluidic MFCs can offer a deeper understanding of certain fundamental processes, e.g., how flow conditions can improve power and current densities.[15,16] However, the adaption of microfluidic MFC platforms is lagging compared with that of macro-systems because the former suffers from lower power outputs, reduced longevity and increased device complexity. Other challenges include flow-induced enhancements to power that fall short of achieving their maximal effect due to competing losses at high flow rates. It is generally believed that shear erosion of the anode-adhered EABs is to blame for this, but a host of competing factors arising from device-level instabilities complicate investigations into this effect. For example, the degradation of gold cathodes by the common oxidizing agent ferricyanide can result in continuous changes to the internal resistance, with high flow reported to exacerbate the problem.[15] The use of highly inert carbon-based electrodes should alleviate this problem while conferring other advantages, such as improved charge transfer kinetics and improved relevance to macro-MFC studies. However, integration of carbon electrodes into microchannels is challenging, so the use of gold electrodes remains a hurdle. Among the most widely acknowledged challenges for microfluidic MFCs is the potential for $O_2$ contamination via diffusion through polydimethylsiloxane (PDMS).[17-19] Despite innovative attempts to address the problem[17-19] it is still standard in microfluidic MFC studies to add $O_2$ scavenging molecules and to use mixed EABs that are more tolerant to $O_2$, but at the expense of complications arising from additional experimental parameters. To achieve reductions in device complexity and cost, a membraneless approach to isolating anolyte and catholyte solutions in laminar-flow chemical fuel cells[20] has been implemented for microfluidic MFCs by selected authors.[21,22] However, its adaption to microfluidic MFCs is not completely straightforward because the EAB causes deviations to flow patterns, thus enhancing the potential for anolyte/catholyte cross-contamination in the opposite compartment.

A general indicator of the remaining challenges for microfluidic MFC developers is the short operating times, most of which are less than two weeks, which is a fraction of the time reported for macro-MFCs. Moreover, short operating times are problems in their own right, as they may not provide an accurate representation of the phenomena under study.[23] Addressing the challenges listed above can improve fundamental studies and can also lead to more efficient microscale applications, such as sensors[24,25] and power supplies for ultra-small electronics or implantable medical devices.[22]

To enable flow studies that are independent of technical complications, we present an easy-to-fabricate membraneless microfluidic MFC with graphite electrodes and a protective barrier against $O_2$ intrusion. Guided by simulations, we optimized the electrode positioning to account for flow deflection around the accumulated biomass at the anode and thus avoid anolyte/catholyte cross-contamination. The result is a robust microfluidic MFC capable of the longest reported operation to date. The system delivers the highest power density outputs recorded to date for any microfluidic MFC with a pure *G. sulfurreducens* EAB, aided by the ability to derive full enhancements to performance from flow.

## Device design and validation

### Fabrication of an oxygen-protected microfluidic MFC with embedded graphite electrodes

Various configurations of membraneless laminar flow fuel cells, including but not limited to MFCs, can be found elsewhere.[20,21,27] Generally, it is practical to fabricate devices with electrodes in a side-by-side configuration using traditional microfabrication techniques because the electrodes are located on one wall. Gold electrodes are usually preferred in microfluidic MFCs due to their compatibility with standard microfabrication techniques,[4,28] particularly as bottom side electrodes in a side-by-side configuration. However, apart from its cost, gold performs sub-optimally as a bioelectrode due to limitations in charge transfer kinetics and in bacterial colonization. Structured electrodes have been implemented to address certain of these drawbacks.[17,21,23,29] However, the instability of gold in the presence of ferricyanide-containing catholyte solutions is often not considered, even though the problem is reported to be worsened at elevated flow rates, thereby limiting one of the main advantages of MFCs in a microfluidic format.[15] The use of gold electrodes in microfluidic MFCs also adds a barrier for transferability to the wider BES community because the vast majority of macro-MFC studies use carbon-based electrodes. The integration of graphite or other carbon-based electrodes is expected to be beneficial because of their excellent charge transfer kinetics,[30,31] favourability to bacterial colonization due to their naturally rough surfaces, and because their inertness should facilitate stable long-term performance.[21,32]

The fabrication process for a strictly anaerobic microfluidic MFC with two side-by-side top-wall graphite electrodes is illustrated in Figure 1a. A photolithographic mould was created (Figure 1a(i)). Two electrodes with dimensions of 10 x 20 mm were cut from a 3 mm-thick graphite plate. Integration of the electrodes into the

PDMS microchannel was adapted from a previously developed fabrication protocol for straight-forward integration of top-wall electrodes.[33] In brief, graphite plates for the anode and cathode were placed on top of the mould channel feature using double-sided tape to ensure good contact. The tape completely surrounded the entire electrode to prevent the previously observed de-activation after immersion in liquid PDMS. The overhanging portion of each electrode was stabilized with a small piece of PDMS (Figure 1a(ii)). A liquid PDMS solution was cured against the assembly, thereby immobilising the electrodes within the device (Figure 1a(iii)). Subsequently, the PDMS device with embedded graphite electrodes was carefully separated from the mould, and the intervening double-sided tape was removed from the portion of the electrode exposed to the channel (Figure 1a (iv)). This method of electrode integration resulted in an exposed electrode surface that was flush with the PDMS wall, thus preventing flow disturbances as liquid flowed across it. Three holes were created for two inlets and one outlet, and the assembly was sonicated in liquid water to remove debris. A 1 M HCl solution was applied to the electrodes using sterile cotton swabs, followed by rinsing of the entire open channel with 70% ethanol and by sterile deionized water. The channel was sealed with a microscope slide (75 × 55 × 1 mm) via standard air plasma activation (Figure 1a(v)). Finally, a protective barrier was added against all exposed PDMS at the top and side surfaces to block gas exchange with the device exterior. This was accomplished using a second glass slide that was plasma-bonded to the top side of the device with the addition of a thin layer of epoxy (LOCTITE®, Ohio, US) along the device edges (Figure 1a(vi)). Access to the fluid connects was achieved by drilling 1/16" holes through the upper glass layer (Figure 4b) with a diamond drill bit. Protection of the inlet and outlet tubing from $O_2$ intrusion was achieved with a gas-barrier coating, as described previously.[9] The final device was fully transparent, protected against small molecule diffusion and capable of generating co-flow of miscible liquids (Figure 1c).

### Protection against small molecule diffusion

Despite the preference for PDMS in most microfluidic MFCs, its use presents a hurdle for anaerobic EABs because of its high permeability to $O_2$ and other small gas-phase molecules such as water,[28,34] which even impacts devices with minimal use of exposed PDMS, such as for spacers.[35] Apart from negatively impacting performance, intruding $O_2$ is expected to become diluted to different concentrations in studies where the flow rate is varied, resulting in complications in the interpretation of results.[19] We first tested the sealing of the microfluidic MFC against small molecule exchange with the ambient conditions outside of the device by monitoring the water evaporation.

Avoidance of water evaporation is important in its own right due to potential side effects, such as increased solute concentrations or precipitation. Two devices were prepared, an unprotected device as shown in Figure 1a(v) and a gas-protected device as shown in Figure 1a(vi). After filling both devices with pure degassed water, the flow was stopped, and the inlets/outlet were sealed shut. Time-lapse imaging was conducted on each device to observe the appearance of air pockets in the channel due to $H_2O$ exchange with the ambient conditions outside of the devices. Figure 2a shows the results from experiments in each device. Bubbles appeared in the unprotected device and continuously grew until 24 hours had elapsed, at which time all water had disappeared from view. Based on the nearly linear slope, it was estimated that 0.16 mm$^3$ of liquid water was lost to diffusion through 1 mm$^2$ of PDMS per day. The effect was slower but still present for thicker devices due to more intervening PDMS. In the second experiment, UV-Vis spectrophotometry was used to accurately observe small changes to the absorbance spectra of the common $O_2$ indicator resazurin after passing through the MFC.[36] In anaerobic aqueous solutions, the initially colourless dye changes colour after it is oxidized to one of two oxidation states at intermediate or near-saturated $O_2$ conditions (see Supporting Information, Section S1, Figure S1) with absorption bands at 573 nm and 605 nm for intermediate and high $O_2$ concentrations, respectively. Figure 2b shows the UV-Vis spectra after passing an anaerobic aqueous resazurin solution (0.15 mg·L$^{-1}$) through the unprotected and gas-protected membraneless MFCs and finally into a sealed glass syringe at the device outlet. After transit through the unprotected MFC, the emerging solution was pink in colour, demonstrating that it had been exposed to $O_2$ within the device, but did not change colour or intensity during the accumulation in the syringe, thus confirming that no further $O_2$ exposure occurred on-chip. The absorbance intensity was observed to have in inverse relationship to total flow rate ($Q_T$) in the range of 1 to 140 mL·h$^{-1}$, suggesting that the $O_2$ concentration does indeed become diluted with increasing $Q_T$. In contrast, Figure 2b confirms that after transit of the indicator solution through the gas-protected device, the accumulated liquid at the outlet remained anaerobic, independent of $Q_T$. Thus, operation at a range of flow rates is possible without affecting on-chip $O_2$ levels.

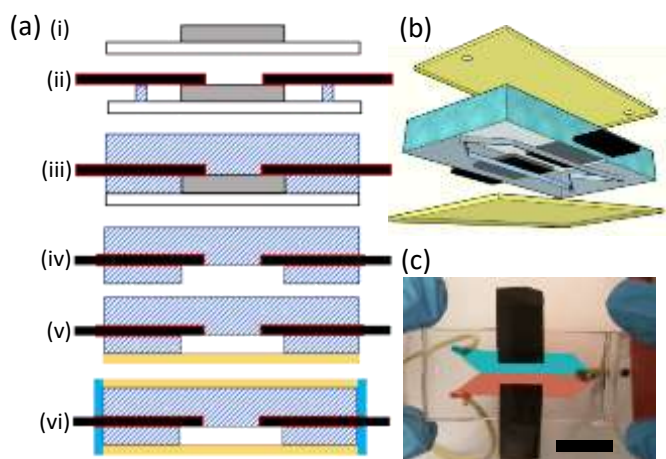

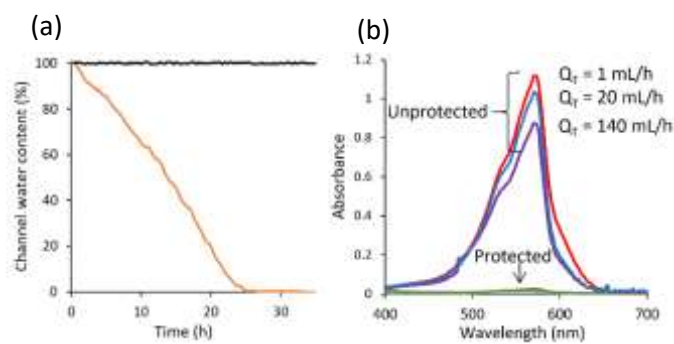

**Figure 1.** (a) Device fabrication steps. (i) Cross-sectional view of the mould with a raised feature that defines the microchannel. The dimensions of the main channel were 160 μm (height), 12 mm (width), and 3 cm (length). (ii) Two 1 cm wide graphite electrodes (black) with a protective external barrier (red) placed on top of the mould channel feature (grey) with small PDMS spacers (blue cross-hatch) for stabilization. (iii) Liquid PDMS (blue cross-hatch) poured over the mould/electrode assembly and cured. (iv) Microfluidic device with embedded electrodes after removal from the mould and removal of the graphite protective barrier in the channel and external to the device. (v) Channel-side (only) sealed with a glass slide (yellow), referred to as the "unprotected" device. (vi) Microfluidic device featuring top and bottom glass sealing and side sealing with epoxy (blue), referred to as the "gas-protected" device. (b) Microfluidic device with drilled inlets and outlet in the top glass sealing layer. Glass and epoxy are shown in yellow and blue. (c) Photo image of the assembled experimental design with blue- and red-coloured liquids to visualise the laminar flows in the device ($Q_{blue} = Q_{red} = 20$ mL·h$^{-1}$). Scale bar is 1 cm.

**Figure 2.** (a) Water content in the microchannel vs. time. Data obtained from a field of view that was approximately 20 percent of the total channel volume in a gas-protected device (black) and unprotected device (orange). (b) UV-visible spectra of aqueous resazurin solutions collected from the outlet of an unprotected MFC at the indicated total flow rates of $Q_T = 140$ (red), 20 (blue) and 1 mL·h$^{-1}$ (purple) and of a gas-protected MFC device at $Q_T = 140$ (grey) and 1 mL·h$^{-1}$ (green).

**Optimization of electrode positioning using flow simulations**

An often-overlooked problem in membraneless MFCs is the possibility of contamination via solution crossover to the opposite electrode compartment due to the presence of an anode-adhered EAB. This can reduce performance due to thermodynamic-based reductions in cell potential and because of ferricyanide toxicity to the EAB. In some work, cross-over contamination has been acknowledged and simulations have shown the result of a diffusive process[17] in an equivalent manner to laminar flow chemical fuel cells. However, this problem has not been strictly evaluated based on the flow deflection around the anode-adhered EAB, which is expected to be significantly more impactful in causing cross-compartment contamination than pure diffusion. This problem is more severe for the popular side-by-side electrode configuration used in this study, due to shorter inter-electrode distances than in opposing wall electrodes. In the Supporting Information, we present simulation results demonstrating several related hydrodynamic phenomena in side-by-side MFCs. These results show that significant displacement of the co-flow interface away from the centre of the channel are a result of the presence of an electrode-adhered biofilm (Supporting Information, Section S2, Figure S2). Visualization of the deflection of the simulated fluid streamlines around the anode-adhered EAB confirms flow deflection as the primary driver of this effect, but secondary flow profiles are also observed, which that can have additional impacts such as mixing at the co-flow interface. Therefore, we begin with a series of simulations to predict the effect of flow on the acetate concentration profiles in the channel, with the goal of minimising cross-contamination.

Figure 3a shows a schematic of the channel used in these simulations. A low aspect ratio of 1:75 (H:W = 0.16 mm:12 mm) was selected to increase the surface-area-to-volume ratio,[21] effectively increasing the contact between the electrodes and the relatively thin fluidic layer above them. The wide channel also allows a large electrode insertion distance ($E_x$) while maintaining sufficient space between the electrodes to accommodate flow deflection around the anode-adhered EAB. The conventional approach is to place electrodes close together to reduce solution resistance,[21] but we hypothesize this can result in electrode contamination from solution crossover in membraneless devices. Thus, we first simulated crossover contamination as the inter-electrode distance was varied (Figure 3a) while equal flow rates of anolyte ($Q_A$) and catholyte ($Q_C$) were applied to each inlet ($Q_A$ = $Q_C$ = 10 mL·h$^{-1}$), with total flow rate $Q_T = Q_A + Q_C$ = 20 mL·h$^{-1}$. To accommodate the widest range of conditions from MFC inoculation to maturity, we analysed the crossover acetate concentration for an EAB thickness ($h_B$) of 10, 30 and 80 μm. Based on our experience and other literature reports, this is a reasonable range, with 80 μm representing a good estimation for mature *G. sulfurreducens* EABs in microchannels. As observed in Figure 3b, electrode separation distances of 4 mm and lower resulted in acetate concentration at the cathode that were more than 10% of the inlet concentration (10 mM) for $h_B$ = 80 μm. For the separation distance of 5 mm, the anolyte concentration at the cathode fell to 0.1 mM, or approximately 1% of the inlet concentration, whereas at 6 mm separation, the concentration decreased to 0.01 mM or approximately 0.1% of the inlet concentration. Because one of the main objectives of this paper was to study flow effects in the absence of device-level instabilities such as crossover contamination, we chose a wide electrode separation distance of 6 mm. Supporting information, Figure S3, shows that simulated flow displacement of the anolyte/catholyte co-flow system and the mixing region at its interface are both increased with $h_B$. It is also noted that total flow rate does not play a significant role in flow deflection (Figure S4). The reader is referred to Supporting Information for a full discussion on these points.

While most experiments use anolyte and catholyte flow rates ($Q_A$ and $Q_C$) that are equal ($Q_A/Q_C$ = 1), simulations are also presented in the Supporting Information for unbalanced flow conditions ($Q_A/Q_C \neq$ 1) which can occur due to user error or maybe implemented intentionally to reposition the co-flow interface within the channel (Figure S5). Finally, supporting simulation of shear stress applied to the anode-adhered EAB were also conducted because this is otherwise difficult to calculate due to the complex flow patterns around the partially occluded channel at the anode position. It is observed that average shear stress against the EAB increased linearly with flow rate and with $h_B$ (Figure S6).

## Results

### Growth of pure-culture *G. sulfurreducens* EABs

We tested the effectiveness of the gas-protected MFC in supporting growth of anode-adhered *G. sulfurreducens* EAB compared with the unprotected device. A pure culture was used as a more rigorous test of the anoxic environment compared with mixed-culture MFCs, which are more tolerant to oxygenated conditions because obligatory and facultative aerobes can produce local anaerobic conditions by consuming $O_2$. The inoculation procedure for both devices was identical and was conducted simultaneously using the same bacteria sub-culture as inoculant. After 2 hours of injecting inoculant at $Q_A$ = $Q_C$ = 0.5 mL·h$^{-1}$, a background potential of 10 mV was recorded across the 25 kΩ external resistor. A 48-hour pre-growth phase followed during which the inoculant was replaced with a growth solution flowing through the channel at the same values of $Q_A$ and $Q_C$. Voltage measurements commenced after the pre-growth phase with both MFCs producing approximately 200 mV. Figure 4a shows the potential after pre-growth (t = 0) immediately after the environmental enclosure was opened, exposing both gas-protected and unprotected devices to ambient (oxygenated) conditions. Exposure of the unprotected device to oxygenated conditions caused the growth in potential to stop, followed by a rapid decrease at 30 h down to 100 mV. This event was followed by a slow decline down to background levels (not shown). Interestingly, the potential in the gas-protected device continued to increase, reaching approximately 500 mV after 2 days and eventually reaching a steady output near 565 mV (see Supporting Information, Section S3, Figure S7 for continuous measurements of potential until the end of the experiment). Figure 4b shows the power density curves obtained at 40 and 80 h for the gas-protected and unprotected devices. We note that the calculated power densities in Figure 4a matched the maximum power density obtained from curves in Figure 4b at the corresponding time. An additional power density curve is shown after 6 months, immediately before the device was sacrificed for SEM (Supporting Information, Section S4, Figure S8). Therefore, the gas-protected design enabled long-term benchtop operation in ambient conditions without the need for $O_2$ scavengers.

### Flow effects on MFC performance

We present the results for the effect of flow on the performance of the gas-protected membraneless MFC with optimized electrode spacing. After growth under $R_{ext}$ = 25 kΩ for approximately 25 days, the MFC reached stable operating conditions. For statistical purposes, power density and polarization curves were acquired weekly during months 2 to 6 while flow was supplied in the range of $Q_T$ = 1 to 40 mL·h$^{-1}$. Figures 5a and 5b show that the maximum power density increased from 376 mW·m⁻² to 430 mW·m⁻² (14%), and the maximum current density increased (at $R_{ext}$ = 10 kΩ) from 905 mA·m⁻² to 1100 mA·m⁻² (22%).

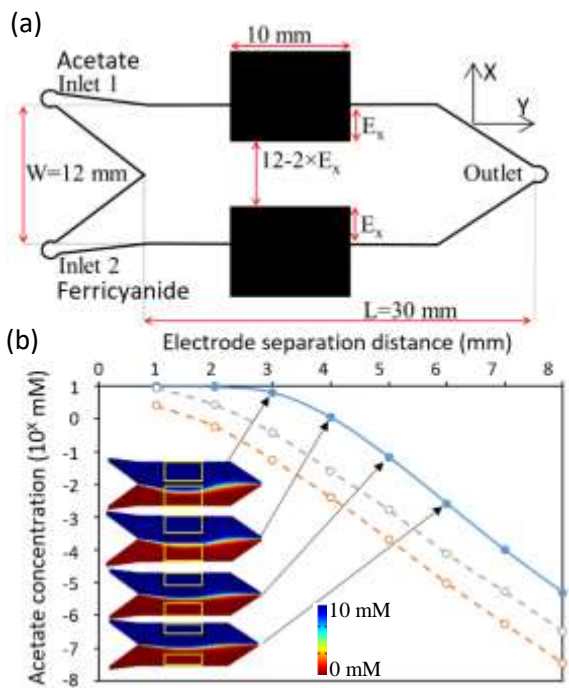

**Figure 3:** (a) Device architecture for flow simulations of the microfluidic MFC with electrode penetration into the channel of $E_x$. Electrode widths and channel dimensions width (W) and length (L) are also shown. Channel height H = 160 μm is not shown. The anolyte (10 mM acetate) and catholyte (30 mM ferricyanide) are supplied at equal flow rates through inlets 1 and 2, respectively, at a total flow rate $Q_T$ = 20 mL·h$^{-1}$. (b) Plot showing the acetate concentration at the cathode edge as electrode separation is varied from 1 to 8 mm for anode-adhered EABs with heights $h_B$ = 80 (blue), 30 (grey) and 10 μm (orange). Inset images show acetate concentration profiles for selected electrode separation distances. Colour bar indicates acetate concentrations.

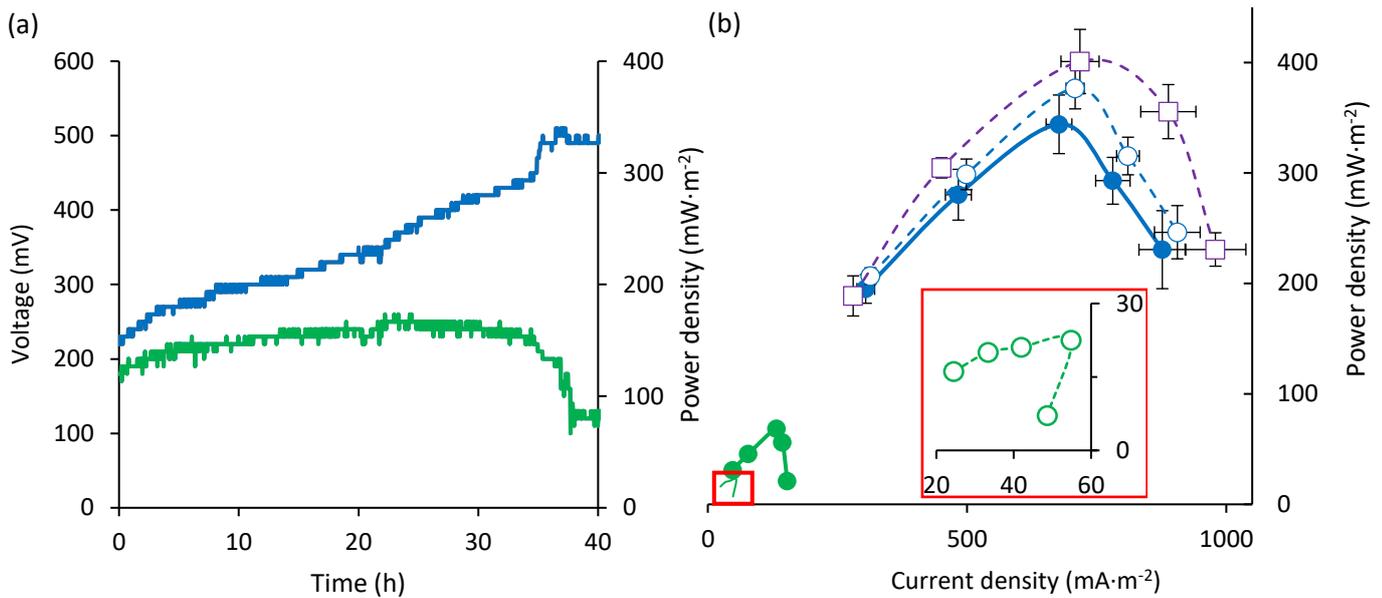

Figure 4. (a) Voltage comparison across the external resistor (25 kΩ) of gas-protected (blue) and unprotected (green) MFCs. (a) Time of 0 h corresponds to the time at which the environmental enclosure was opened after 48 h pre-growth under a 10 mM acetate growth solution. The second vertical axis shows the calculated power density. (b) Power density curves for both devices at t = 40 h (solid) and t = 80 h (dashed) after opening the anaerobic enclosure for the gas-protected (blue) and unprotected (green) MFCs using external resistances $R_{ext}$ = 10, 16, 25, 40, and 70 kΩ. A third power density curve is shown for the gas-protected device at t = 6 months (dashed purple). Error bands in (b) are calculated from 3 separate measurements but are smaller than the data markers for the unprotected device. The inset (red box) shows a zoom view of the highlighted curve for the unprotected device after 80 h. In all cases, the flow rate was $Q_T$ = 1 mL·h⁻¹.

We note that the maximum power density was always observed when the external resistance was set to $R_{ext}$ = 25 kΩ. Therefore, flow did not significantly affect the internal resistance in the flow rate range of 1 to 40 mL·h⁻¹. Higher rates were not attainable while acquiring power density and polarization curves because the length of time required for the measurement surpassed the discharge time of the syringes used. To extend the range of applicable flow rates, voltage was measured across a single external resister of $R_{ext}$ = 25 kΩ and converted to power density using Eqn. 3, as discussed in the experimental section. Figure 5c shows the changes to the potential and calculated power density as the total flow rate was changed between a baseline value (1 mL·h⁻¹) and an elevated value (between $Q_T$ = 2 and 140 mL·h⁻¹) with a duty cycle of approximately 15 min. Figure 5d shows that both methods produced a similar response to flow rate from 1 to 40 mL·h⁻¹ within the experimental error, as indicated by the error bars. The constant resistor method produced changes in power density from 368 mW·m⁻² at $Q_T$ = 1 mL·h⁻¹ to 456 mW·m⁻² at $Q_T$ = 140 mL·h⁻¹ with no further increases for flow rates as high as 200 mL·h⁻¹ (data not shown). This is equivalent to an increase of 24% over baseline flow rates ($Q_T$ = 1 mL·h⁻¹) (Figure 5d inset). An important corollary is that the benefits of enhanced flow attain their maximum effect on power density before the onset of sheer-induced erosion.

The growth in current density at 25 kΩ was nearly 8% in the same range of $Q_T$ values (Figure 5d), but separate measurements with $R_{ext}$ changed to 10 kΩ showed values ranging from 925 to nearly 1100 mA·m⁻² or approximately 19% growth. These improvements can be ascribed to enhanced metabolic activity, largely due to enhanced nutrient availability. However, other factors could have contributed. For example, although EAB de-acidification was not observed in previous chronoamperometry studies of the same EAB type exposed to 10 mM acetate concentration,[9] it could have played a role here due to the combination of higher flow rates, thinner EAB and the lack of an external driving potential.

## Discussion

Next, we discuss the results presented above with comparisons to the literature. Table 1 summarizes the literature results from other studies using pure cultures of *G. sulfurreducens* EAB in microfluidic MFCs. Included are MFCs with and without membranes to demonstrate that until now, membraneless MFCs have underperformed compared with membrane MFCs, likely due to uncontrolled cross-contamination. In this work, three design features were combined: (i) $O_2$ elimination without the use of a scavenger, (ii) avoidance of cross-contamination, and (iii) use of graphite electrodes. Implementation of these features together on the same device resulted in the highest power and current densities and the longest operating times to date. Table 1 does not include results from mixed-culture or *Geobacter*-enriched EABs, which are generally preferred in the literature due to improved cell potentials related to better tolerance to oxygen. Even among this wider cross-section of the literature, the longest operational times were just over 15 days,[17,19] although one notable exception is a membraneless microbial electrolysis cell containing pure *G. sulfurreducens* with over 30 days of operation achieved by housing an unprotected PDMS device in a home-built acrylic anaerobic chamber without the use of ferricyanide.[24] Therefore, it appears that the exclusion of ferricyanide might indeed improve gold electrode longevity, whereas the inert graphite electrodes used in this study could still benefit from fast ferricyanide reduction kinetics. We note in passing that L-cysteine is known to form monolayers on gold surfaces,[37] which does not appear to limit the performance of gold electrode MFCs on a short timescale,[19] but it might over longer timescales. In any case, although graphite electrodes should avoid this problem, no $O_2$ scavenging was used in this work.

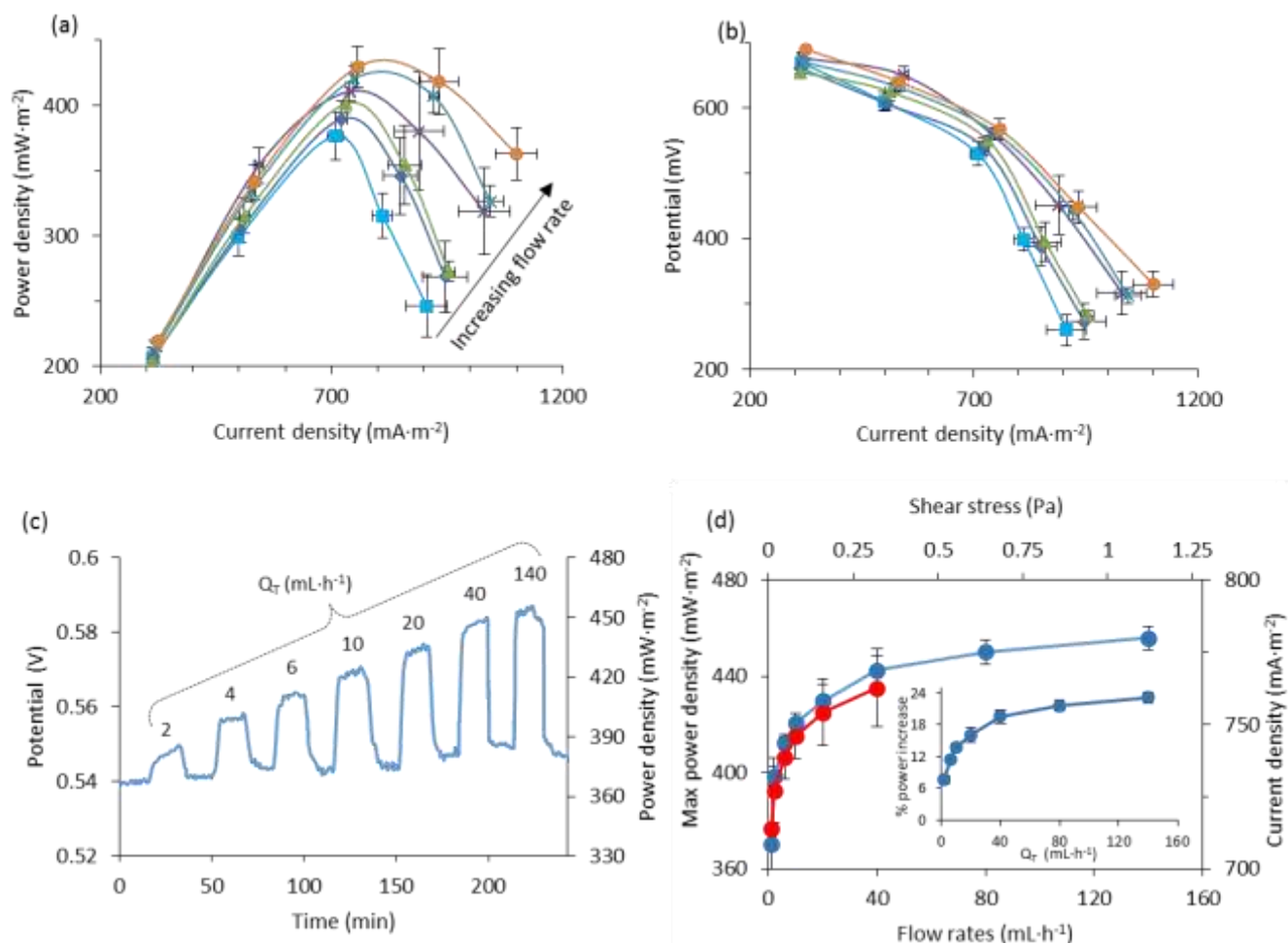

Figure 5. Performance of the gas-protected microfluidic MFC with 6 mm electrode spacing. (a) Power density curves and (b) polarization curves at total flow rates $Q_T = 1$, 2, 6, 10, 20 and 40 mL·h[-1]. (c) Results of pulsing $Q_T$ on the measured potential across a 25 kΩ resistor and the corresponding power density outputs for 7 elevated flow rates separated by exposure to a baseline flow rate ($Q_T = 1$ mL·h[-1]). (d) Maximum power density from data in (a) (red) and power density from (c) (blue). Inset shows percent increase in power ($\Delta p/p_{static}$ x 100%) with $Q_T$. In all cases, error bars are the result of 5 separate measurements collected during months 2-6, inclusive.

Despite the demonstrated advantages of our system, considerable room for improvement exists considering that the highest performing microfluidic MFCs (0.6, 1.8 and 2.5 W·m[-2]) were also membraneless configurations with graphite electrodes and gas-impermeable plastic materials.[18,35,38] These studies benefited from the lowest internal resistances reported in the literature (ca. 1 to 2 kΩ) due to short inter-electrode distances (2 to < 1 mm) and thick EABs (hundreds of microns) and could also benefit from flow, albeit over a small range of flow rates and with the use of mixed species EABs. However, cross-contamination was hypothesized to limit the outputs, with reductions measured at elevated flow rates, whereas the device presented in this work reached maximum stable power levels without losses at high flow rates. Therefore, it is possible that further optimization can lead to better performance. For example, our goal of limiting contamination to 0.05 mM acetate could be relaxed or even maintained with closer electrodes by adjusting $Q_A/Q_C < 1$. Other avenues to reducing internal resistance include long-term culture and growth under elevated flow rates to produce compact and conductive EABs.

## Conclusion

The goal of this paper was to implement design enhancements that could produce more consistent performance of membraneless microfluidic MFCs for greater impact among the wider BES community. To this end, we developed an easy-to-fabricate device with optimized placement of graphite electrodes and $O_2$ protection. The realization of this device revealed several advantages, and principle among them was the period of continuous operation, which we believe is an important general figure of merit that reflects the device functionality. In this work, we achieved 6 months of continuous operation before sacrificing the device for SEM, the longest duration yet reported. This duration was achieved using an EAB from pure *G. sulfurreducens*, without the inclusion of an oxygen scavenger. The device stability under flow also appears to have been critical to achieving continuous flow-induced improvements to power outputs without

**Table 1:**

| Membrane | µMFC material | Anode material | Cathode material | $P_{max}$ (mW·m$^{-2}$) | $I_{max}$ (mA·m$^{-2}$) | $R_{in}$ Ω | Distance µm | Operation time | Ref |
|---|---|---|---|---|---|---|---|---|---|
| None | PDMS | Gold | Gold | - | 18 | - | 200 | - | 39 |
| None | PMMA | Gold | Gold | - | 65 | - | 100 | 7 days | 40 |
| Nafion 117 | PDMS | Si/CNT | Gold | 36 | 76 | 25 K | - | - | 41 |
| Nafion 212 | PMMA | Gold | Gold | 120 | 600 | 200 K | - | 10 days | 42 |
| None | Glass/ PDMS | Graphite | Graphite | 456 | 1100 | 25 K | 6000 | 6 Months | Our work |

Dashes indicate non-reported information losses at flow rates as high at 200 mL·h$^{-1}$, corresponding to shear forces of more than 1.6 Pa. With assurances that flow-based changes to power output are uncomplicated by device-level instabilities, studies into the critical shear stress required for biofilm erosion can now be undertaken based on the biofilm type or growth history. In addition, other mechanistic studies into molecular transport and conversion kinetics will be facilitated. The next round of technical advancements should include optimization of the operating conditions and device improvements that target higher performance. These improvements can include manipulation of the co-flow interface position to resist the natural tendency of displacement towards the cathode. Closer electrode placement can further reduce internal resistance. Another approach is the introduction of a third inlet for an electrolyte stream with low solution resistance to separate the anolyte and catholyte streams without introducing added internal resistance, as demonstrated previously in a microfluidic chemical fuel cell.[43] Finally, the device should be tested with mixed-species biofilms to expand the range of applicable substrate molecules and to provide better connection with typical macro-MFC studies.

## Experimental Section

### Microfabrication tools and techniques

Moulds for casting PDMS microchannels were created via photolithography. The microchannel motif was designed using computer aided design software (DraftSightTM, Dassault Systemes, France), and the photomask was printed using a photoplotter (FPS25000, Fortex Engineering Ltd., UK). The photomask was subsequently placed on top of a photoresist layer (SY300 film, Fortex Engineering Ltd., UK) with 160 µm thickness that was previously adhered to a 75 x 50 x 1 mm glass slide (12-550C, Fisher Scientific, Canada) using a dry film laminator (FL-0304-01, Fortex Engineering Ltd., UK). The photoresist was selectively cross-linked through the mask with a UV exposure system (AY-315, Fortex Engineering Ltd., UK), and excess photoresist was removed using developer and rinse solutions (SY300 Developer/Rinse, Fortex Engineering Ltd., UK). Microfluidic devices were cast against the resulting mould using liquid PDMS and cross-linker (Sylgard184, Dow Corning, Canada) at a 10 to 1 ratio, followed by curing at 70ºC for 4 hours. Inlet and outlet holes were punched through the device, and glass bonding to the PDMS was achieved using air plasma (PCD-001 Harrick Plasma, Ithaca, USA) for 90 seconds at 650 mTorr. Graphite electrodes (GraphiteStore.com Inc., USA) were wrapped in a tight-fitting adhesive barrier to prevent electrode deactivation, which was otherwise observed after contact with liquid PDMS.

### *Geobacter sulfurreducens* preparation and device start-up

All air-sensitive operations were performed in an anaerobic glove box (10% H$_2$, 10% CO$_2$ and 80% N$_2$). *G. sulfurreducens* (wild type, strain PCA, ATCC 51573) were thawed from ceramic beads stored at -80°C and cultured in a liquid nutrient medium containing fumarate. A maximum of 8 sub-cultures could be generated from one culture. Sub-cultures were used as inoculant solutions that included 40 mM sodium fumarate (as an electron acceptor) to promote growth without the presence of an anode. In addition, the nutrient medium consisted of a sodium acetate carbon source (CH$_3$COONa (10 mM), NaH$_2$PHO$_4$ (3.8 mM), NaHCO$_3$ (30 mM), KCl (1.3 mM), NH$_4$Cl (28 mM)), a trace mineral supplement (10 mL·L$^{-1}$) (ATCC® MD-TMS™), and a vitamin supplement (10 mL·L$^{-1}$) (ATCC® MD-TMS™). Before use, all nutrient media were sterilized by autoclaving at 110°C and 20 psi for 20 min. Because sodium fumarate could not be autoclaved, it was added afterwards through a sterilized filter in the glovebox. A second identical solution without fumarate was also prepared as the main nutrient source for use after inoculation to promote growth on the anode. No O$_2$ scavenger molecules were included in any solutions used in this study.

The MFC anode was inoculated with the sub-culture under a flow rate of 0.5 mL·h$^{-1}$ while a catholyte (30 mM ferricyanide solution in a pH 7.0 sodium phosphate buffer) was co-flowed at 0.5 mL·h$^{-1}$ for 2 h via separate syringe pumps (PhD 2000, Harvard Apparatus, MA). During inoculation, the MFC anode and cathode were connected through an external resistor $R_{ext}$ = 25 kΩ. The inoculum was replaced with a sterilized fumarate-free nutrient medium, and the MFC was allowed to undergo a pre-growth phase for 48 h under the same flow conditions. As a control, all inoculation and pre-growth processes were conducted in a sealed anaerobic enclosure (McIntosh and Filde's, 28029, Sigma-Aldrich) filled with an anaerobic gas (20% CO$_2$ and 80% N$_2$). All reactors were operated in a temperature-controlled laboratory (23 ± 0.5°C). Following inoculation, the sub-culture inoculum solution was replaced with the fumarate-free nutrient solution such that only *G. sulfurreducens* that were attached to the anode could complete the Kreb's cycle via anode respiration. Experiments were conducted for 6 months until the device was sacrificed for SEM imaging of the anode-adhered EAB.

**Electrochemical measurements**

All electrochemical measurements were collected during a continuous feed of anolyte and catholyte through the microfluidic MFC. A microcontroller data acquisition system (Atmega328pMmicrochip Technologies, Arizona, USA) was used to measure the voltage V at intervals of every 10 min across an external resistor $R_{ext}$ ($\Omega$) connecting the anode and the cathode. The potential drop measurements could be transformed into current measurements using Ohm's law:

$$I = V/R_{ext} \qquad \text{(Eqn. 1)}$$

The power density was measured via two methods. The first used a commercial potentiostat (VersaSTAT 4, Princeton Applied Research, Oak Ridge, USA) that automatically switched $R_{ext}$ values while measuring the voltage drop to obtain polarization curves using Eqn. 1 and power density curves using Eqn. 2:

$$P = V^2/R_{ext} \qquad \text{(Eqn.2)}$$

The maximum power with this technique is obtained when $R_{ext}$ is equivalent to the internal resistance ($R_{int}$). Due to the long duration of this experiment, a faster second approach was used in which the potential drop was continuously measured across a single external resistor (with $R_{ext} = R_{int}$), and power was calculated using Eqn. 2. Each week, the power curves were acquired via the first method to measure the approximate $R_{int}$ so that $R_{ext}$ could be properly set ($R_{ext} = R_{int}$). Measurements of power density ($P_{density}$) were obtained by dividing P from Eqn. 2 by the exposed electrode area in the channel, A = 30 mm$^2$:

$$P_{density} = P/A \qquad \text{(Eqn. 3)}$$

**UV-Vis measurements of O$_2$ indicator resazurin**

The presence of O$_2$ in the microfluidic MFC was measured using a standard 0.15 mg·L$^{-1}$ aqueous solution of the indicator resazurin.[26] Each device was injected with the resazurin solution using two air-tight glass syringes connected to the inlets, and samples were collected using one glass syringe at the outlet. Approximately 2-3 mL of the resazurin solution were collected from the outlet, and absorbance spectra were measured via UV-Vis spectrophotometry (Shimadzu, UV-2450 spectrophotometer, Japan) for analysis of peak position and intensity. An O$_2$-purged glovebox was used during the transfer to an air-tight UV-Vis cuvette. All experiments and measurements were conducted at 23 ± 0.5°C.

**Imaging**

Optical microscopy was used to observe H$_2$O evaporation through the MFC walls based on the appearance of air bubbles after the flow was stopped. The air volume of the air space in the channel was quantified by image analysis using ImageJ v1.51 (National Institute of Health, USA). Scanning electron microscopy (SEM) analysis was performed at the end of experiment to confirm *Geobacter* attachment to the anode and to visualize and measure the EAB height. The anode-adhered EAB was fixed by washing with 2.5% glutaraldehyde in a phosphate buffer solution (flow rate of Q=0.5 mL·h$^{-1}$) for 2 h before the electrode was removed from the device. After electrode removal, the EAB was placed in the fixation solution overnight and subsequently transferred to a solution with 1% osmium tetroxide for 1.5 h before a final washing with pure PBS buffer (pH 7.4). The sample was dehydrated in a series of aqueous ethanol solutions for 15 min each, starting with equal parts ethanol and water and ending with 100% ethanol. A thin layer of gold metal was applied on dried samples using a sputter coater (Nanotech, SEM PREP 2) prior to SEM imaging (JEOL JSM-6360 LV).

**Flow simulations**

Computational studies of the anolyte and catholyte co-flow properties at a variety of flow rates were accomplished using three-dimensional simulations (COMSOL Multiphysics 4.2a, Stockholm, Sweden). A fine mesh was used with physics for laminar flow and transport of dilute species in an incompressible liquid. The molecular diffusion values applied for ferrocyanide and acetate molecules were $D_{Fe} = 1.2 \times 10^{-5}$ cm$^2$·s$^{-1}$ and $D_{Ac} = 7 \times 10^{-6}$ cm$^2$·s$^{-1}$, respectively.

## Acknowledgements

The authors thank Luc Trudel and Laurent Smith (U. Laval) for their help with culture of *G. sulfurreducens* and Richard Janvier (U. Laval) for technical assistance with SEM. This research was supported by generous funds from the Natural Sciences and Engineering Research Council of Canada, the Canadian Foundation for Innovation and Sentinel North. Greener is the recipient of an early researcher award from the Fonds de Recherche du Quebec – Nature et Technologies. Greener (high risk, high reward) to study microbiological systems using microfluidics. The authors also thank Molly K. Gregas for assistance with technical edits.

**Keywords:** Microbial fuel cell • microfluidics • computational fluid dynamics • graphite electrodes • *Geobacter sulfurreducens*

Insert graphic for Table of Contents here.

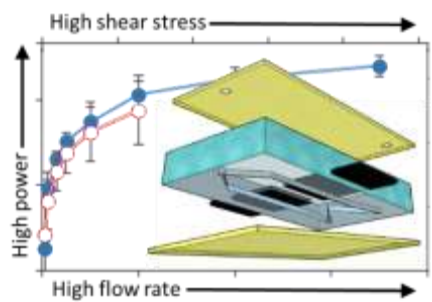

Supporting Information:

# A high performance membraneless microfluidic microbial fuel cell for stable, long-term benchtop operation under strong flow

Mehran Abbaszadeh Amirdehi,[a] Nastaran Khodaparastasgarabad,[a] Hamza Landari,[b] Mir Pouyan Zarabadi,[a] Amine Miled,[b] Jesse Greener [a,c] *

[1] Département de Chimie, Faculté des sciences et de génie, Université Laval, Québec City, QC, Canada.

[2] Département de Génie électrique, Faculté des sciences et de génie, Université Laval, Québec, Canada

[3] CHU de Québec, centre de recherche, Université Laval, 10 rue de l'Espinay, Québec, QC, Canada.

* Corresponding author: jesse.greener@chm.ulaval.ca

**Sections:**

S1. Resazurin $O_2$ indicator

S2. Simulations

S3. Long-term voltage measurements

S4. Scanning electron microscopy

## Section S1-Resazurin O₂ indicator

Figure S1 shows the colour response of a resazurin $O_2$ indicator solution in bulk by eye and by spectrophotometry as it passes from purely anaerobic conditions (colourless) to mono- (pink) and di-oxygenated (blue) states during exposure to ambient air conditions. Resazurin has one of the highest Kreft's dichromaticity index,[1] meaning the perceived change in colour hue is high, making it very sensitive to low $O_2$ concentrations. For example, the onset of pink occurs at pseudo-anaerobic conditions. The main paper presents results using the same solution to verify the suitability of the $O_2$ protection barrier for the anaerobic microfluidic MFC demonstrated in this work.

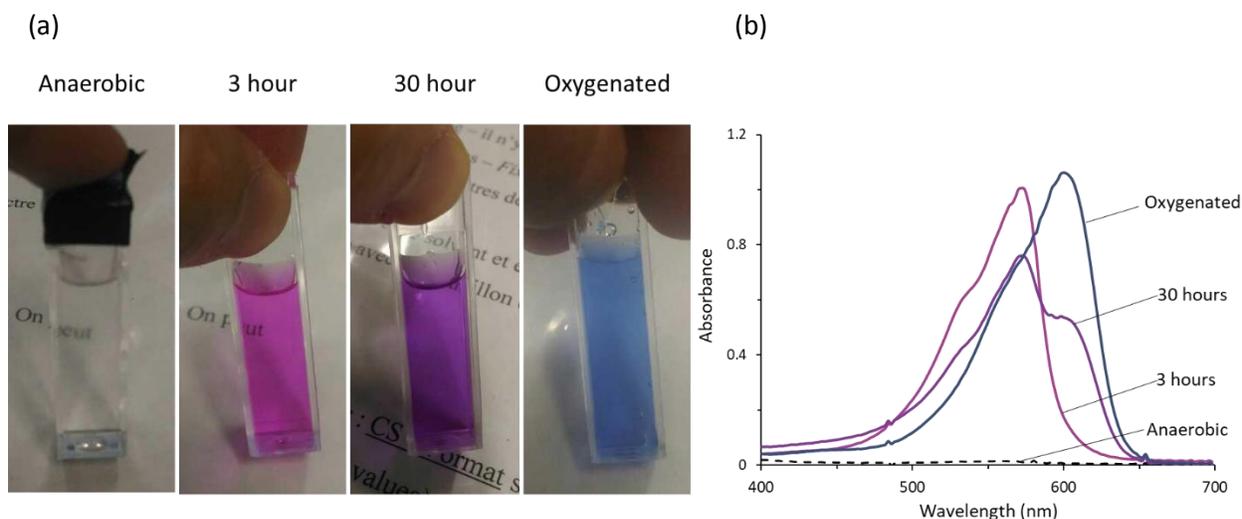

**Figure S1**. (a) Qualitative changes to resazurin solution under indicated exposure times to ambient (oxygenated) air conditions. (b) Corresponding UV-Vis spectra.

## Section S2-Simulations

This section shows a series of hydrodynamic simulations using Comsol Multiphysics with physics for flow of incompressible fluid flow and diffusion mass transport of dilute solutes.

Figure S2 shows results from a simulation within the microfluidic MFC used in this study featuring an EAB with height $h_B = 80$ µm (see schematic of the channel cross-section, Figure S2a) and the resulting displacement of the anolyte toward the cathode chamber (Figure S2b). It is clear from Figure S2c that deflection of flow streamlines are primarily responsible for the co-flow interface position. Figure S2d shows the z-component of flow velocity ($v_z$), which is mostly zero throughout the channel except near the anode-adhered EAB. This can be due to a combination of Dean vortices and pressure imbalances. While small, approximately 1000 times less than the downstream component of velocity ($v_y$), this could have an important impact on mixing especially for electrode that are closely spaced.

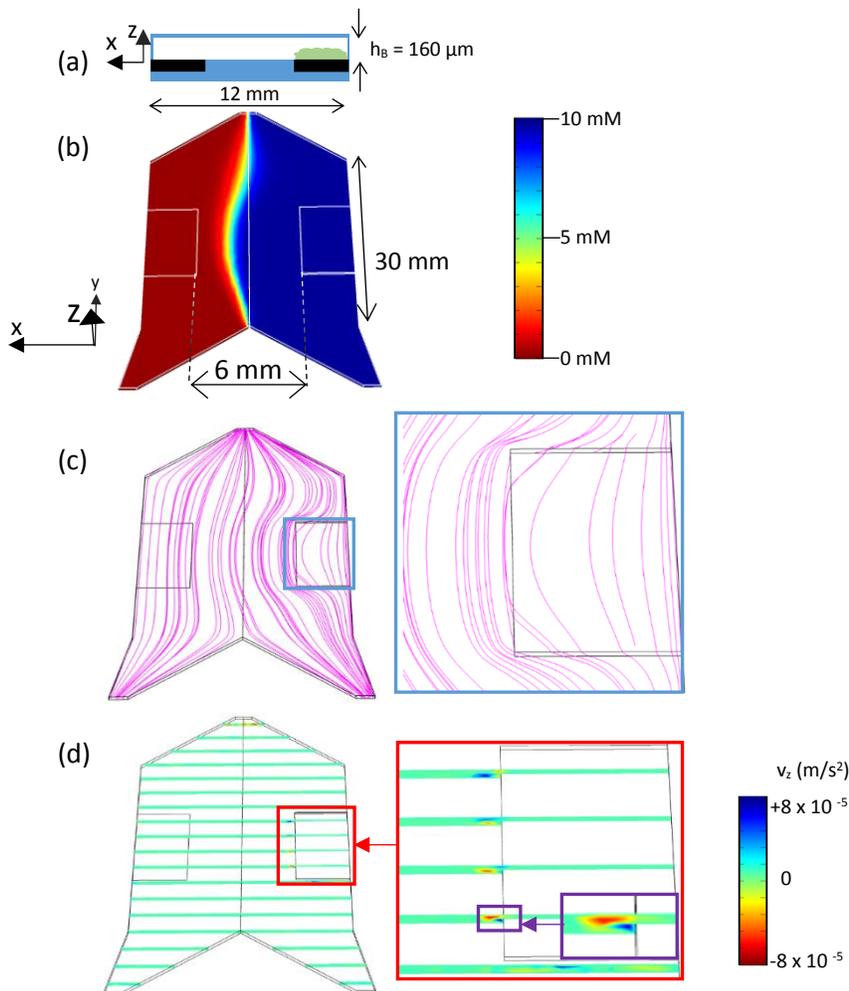

**Figure S2**. Simulations of chemical and hydrodynamic profiles in a microfluidic MFC with the same design as was used in the experimental work. (a) Schematic showing the device cross-section with embedded electrodes and an anode-adhered electroactive biofilm protruding from the anode surface. (b) Concentration profile of a 10 mM acetate anolyte solution (blue) imposed against a catholyte solution (red) consisting of 0 mM acetate and 30 mM ferricyanide catholyte solution (not shown). (c) Fluid streamlines showing deflection around the electrode adhered EAB, with zoom view in the inset of the blue highlighted area in the main figure. (d) Secondary flow velocity profiles in the z-direction at the edge of the bio-anode, with two inset zoom views from red and purple highlighted zones. Z-component velocities are blue (negative direction) red (positive direction) near the electrode, and zero (green) elsewhere. Simulation conditions in (b)-(d) were the same in all cases, anode-adhered electroactive biofilm thickness, $h_B = 80$ μm, and flow rate, $Q_T = 20$ mL/h, with $Q_A/Q_C = 1$. Electrodes were placed at the channel mid-position with a down-stream length of 1 cm and anode-cathode separation of 6 mm. Channel dimensions are shown in (b).

Simulations on the position and width of the anolyte/catholyte co-flow interface were conducted under balanced flow conditions ($Q_A/Q_C = 1$) with variable $h_B$. Based on the literature values, we repeated the simulations, anode-adhered EABs ranged from 10 to 80 μm (Figure S3). As expected, increases to $h_B$ tend to push the co-flow position closer to the cathode. It is observed that the mixing region at the interface increased with $h_B$. This is somewhat counter intuitive because the local fluid velocity should increase with $h_B$ because of local decreases to the channel volume, reducing liquid transit times and diffusion distances. However, as discussed above, mass transport perpendicular to the flow channel, is not strictly diffusion limited, but includes secondary flow profiles in the z and x directions, including the likelihood for vortices near the EAB/liquid interface. Though not completely analogous with the present case where we include the presence of an anode-adhered EAB, this result agrees with previous simulations showing that structured electrodes in a membraneless device.[2] In short, the thickness of the anode-adhered EAB must be accounted for in simulations of the position of both the co-flow interface, and its width.

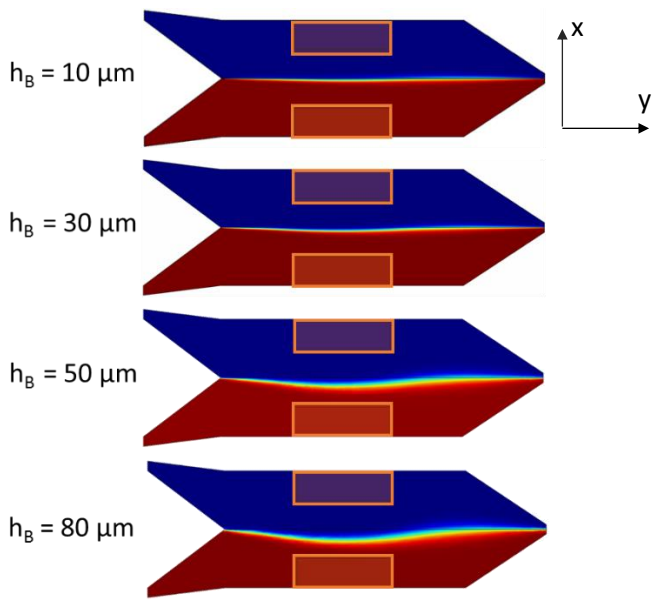

**Figure S3**. Effect of $h_B$ on concentration gradient and position under balanced flow of the anolyte (blue) and catholyte (red) under balanced flow conditions, $Q_A/Q_C=1$ and $Q_T = 20$ mL·h$^{-1}$.

Figure S4 shows the x-y contours for acetate concentrations of 5 mM and 0.05 mM. These two concentrations represent the mid-point between the two solution streams and the lowest acceptable crossover concentration, respectively. As expected, the displacement of the co-flow interface was less pronounced for $h_B = 10$ µm (Figure S4a) than for $h_B = 80$ µm (Figure S4b). The co-flow mid-point could be displaced by up to 1 mm (for $h_B = 80$ µm), whereas the low-concentration threshold contour could extend an additional 1 to 2.5 mm further (for $h_B = 10$ and 80 µm, respectively). In addition, each experiment was conducted at two extreme flow rates of $Q_T = 1$ and 140 mL·h$^{-1}$. Lower flow rates resulted in slightly pronounced displacement distances, likely due to longer diffusion times, although subtle effects related to secondary flow could also have played a role.

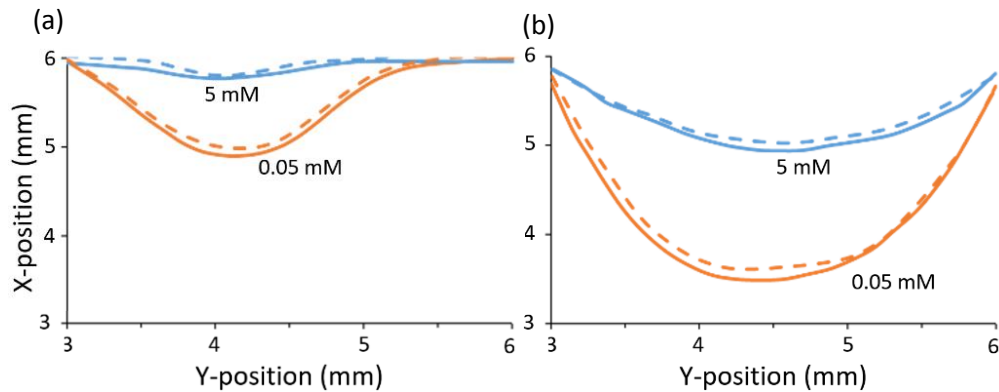

**Figure S4.** Displacement of the co-flow interface for membraneless MFCs around an electrode-adhered EAB (edge at 6 mm) with EAB height $h_B$ = 10 μm (a) and $h_B$ = 80 μm (b). Acetate concentration contours are shown for 5 mM (blue) and 0.05 mM (orange) at total flow rates of 1 and 140 mL·h$^{-1}$, with both inlet flow rates held equal. The x- and y-coordinates on the plot axes correspond to the directions perpendicular and co-linear to the channel length, as shown in Fig S3a.

In Figure 5a we consider the effect of unbalanced flow ($Q_A/Q_C \neq 1$) to determine the acceptable range of flow rate ratios $Q_A/Q_C$ or $Q_C/Q_A$, for which the positioning of certain threshold contours was directly adjacent to (but not touching) the cathode (Figure S5b) or bio-anode (Figure S5c), respectively, as $h_B$ increased from 10 to 80 μm. For example, for $Q_A$ = 20 mL·h$^{-1}$ and $Q_C$ = 10 mL·h$^{-1}$ (flow rate ratio of $Q_A/Q_C$ = 2) placed the the 0.05 mM acetate (0.5% of the inlet concentration) the position in Figure S5b for $h_B$ = 10 μm. This position remained unchanged independent of total flow rates as long as flow rate ratio and $h_B$ were constant. As $h_B$ increased, so too did the propensity for flow deflection around the anode-adhered EAB, requiring progressively lower values of $Q_A/Q_C$ to maintain the 0.05 mM acetate concentration in the same position. Other 0.05 mM contour lines, could be positioned in the same place as shown in Figure 5b for any any $h_B$ value by adjusting the flow rate ratio as shown in Figure S5a (open blue symbols). Higher concentrations of contaminating anolyte at the cathode occurred for larger flow imbalances (larger values of $Q_A/Q_C$) but followed the same trend. For example, Figure 5a (open green symbols) shows $Q_A/Q_C$ values required to position the 5 mM acetate concentration contour (50% of the inlet concentration) at the cathode. Contamination of the anode by the catholyte became more difficult as $h_B$ values increased due to the natural tendency to deflect the flow streamlines away from that side of the device (Figure 5a, solid symbols). Unlike the acetate concentration contours adjacent to the cathode, those adjacent to the anode were strongly dependent on $h_B$, showing a tendency to "surround" the anode for large $h_B$ values due to the tendency of the anolyte stream to cause local deflections of streamlines away from the channel wall (Figure 5c). This effect was not observed for low $h_B$, similar to concentration profile in Figure S5b, because of the lack of EAB is present. The shaded parts of the curves in Figure 5a show the range of flow rate ratios that could achieve confinement of the specified acetate and ferricyanide concentration thresholds (expressed as a percentage of inlet concentration) between the cathode and anode.

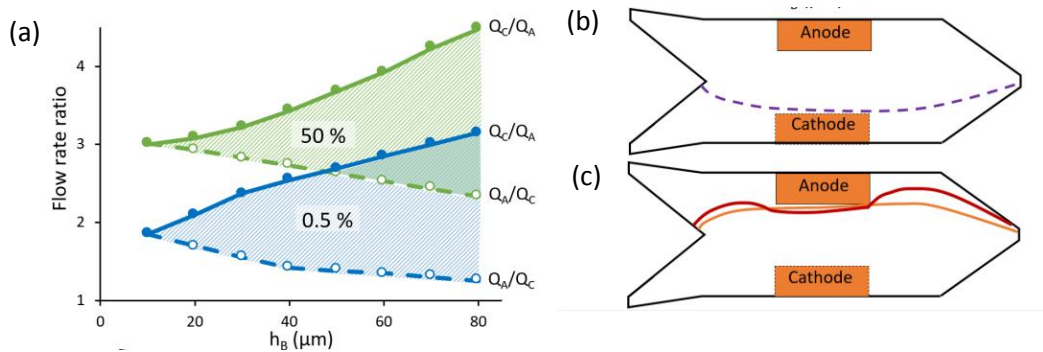

**Figure S5**. (a) Flow rate ratios required for positioning of 0.05 mM acetate (0.5% of inlet concentration, blue) and 5 mM acetate (50% of inlet concentration, green) contour lines near the cathode (dashed lines, open circles) and the 0.15 mM ferricyanide (0.5% of inlet concentration, blue) and the 15 mM ferricyanide (50% of inlet concentration, green) contour lines adjacent to the anode (solid lines, filled circles). The definition of the flow rate ratio is indicated for cathode- and anode-adjacent concentration profiles. Typical cathode- (b) and anode-adjacent (c) concentration contours. (b) is valid for all $h_B$ values. (c) Anode-adjacent contours for (i) $h_B = 10$ μm and $Q_C/Q_A = 2$ (orange) and (ii) $h_B = 80$ μm and $Q_C/Q_A = 3.3$ (red).

Finally, computer simulations results show that the average shear stress across the top of the EAB showed linear increases with total flow rate ($Q_T$) and with $h_B$, in Figure S6a and S6b, respectively. As this is the first results, to our knowledge, which include the role of $h_B$ on the shear stress values, we include for comparison, results for $h_B = 0$, the effective conditions modeled where biofilm height is not included in the simulation.[2] In Figure S6b, we see that this trend is also linear with $h_B$, when keeping $Q_T$ constant.

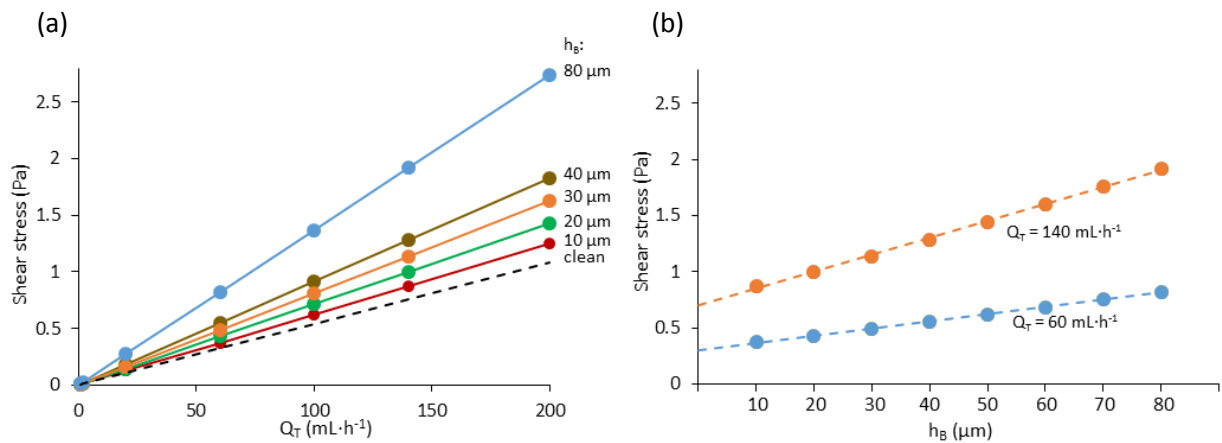

**Figure S6**. Effect of flow rate and EAB height on shear stress. (a) Changes in shear stress as a function of total flow rate ($Q_T$) for EAB heights $h_B = 10, 20, 30, 40, 80$ μm. Dashed line shows results for no EAB ($h_B = 0$). Linear fits are shown (solid lines) with $R^2 = 1$ in all cases. (b) Trends in shear stress as a function of $h_B$ for two selected total flow rates $Q_T = 60$ and $140$ mL·h⁻¹. Linear fit is shown (dashed lines) with $R^2 = 0.998$. Flow rate ratio was $Q_A/Q_C = 1$ in all cases.

## Section S3-Long-term voltage measurements

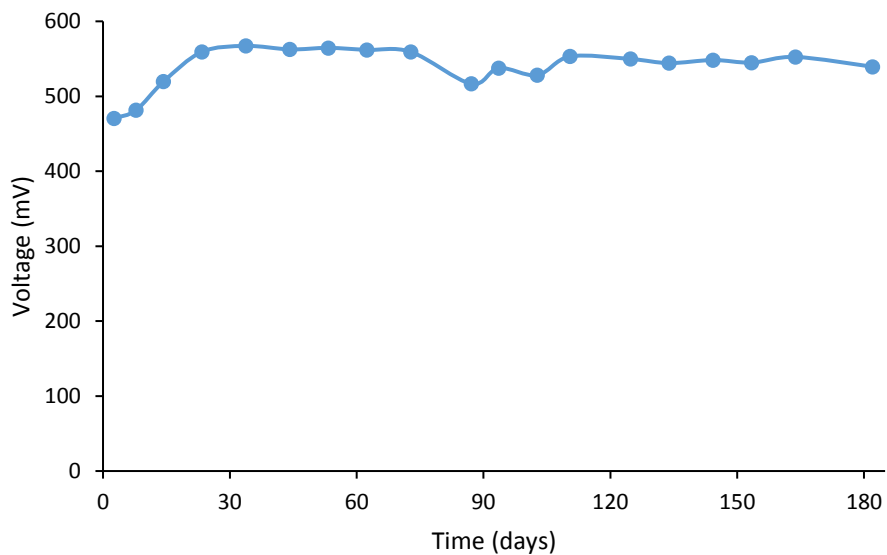

**Figure S7**. Potential measurements across $R_{ext}$ = 25 kΩ following the end of Figure 6a (main paper). Here, t = 0 represents 40 h after exposing the MFC to ambient conditions.

## Section S4-Scanning electron microscopy

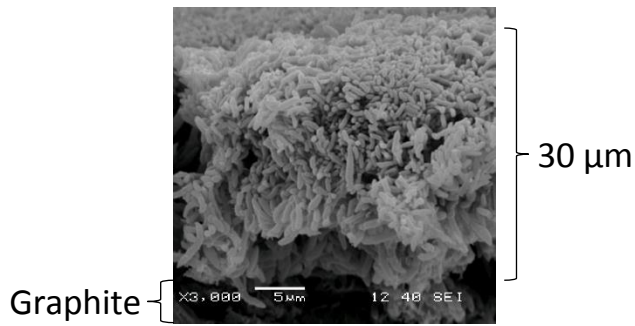

**Figure S8**. SEM of an anode-adhered EAB.

## Section S4-References